# A Novel Channel Coding for Progressive Transmission of Medical Images

P. Jagatheeswari, Dr. M. Rajaram

**Abstract**—A novel channel coding scheme for progressive transmission of large images is proposed. The transmission time, low distortion reconstructed image and low complexity are most concerned in this paper. In the case of medical data transmission, it is vital to keep the distortion level under control as in most of the cases certain clinically important regions have to be transmitted without any visible error. The proposed system significantly reduces the transmission time and error. The progressive transmission is based on the process that the input image is decomposed into many subblocks each to be coded, compressed, and transmitted individually. Therefore, firstly the image is segmented into a number of subblocks and then the discrete wavelet transform decomposes each subblock into different time-frequency components. Finally the components are coded for error control and transmitted. The complete system is coded in VHDL. In the proposed system, we choose a 3-level Haar wavelet transform to perform the wavelet transform for each subblock. It is simple, faster and easier to implement when compared with other transform method. The channel coding used here is Hamming code which is a simpler and efficient forward error control code.

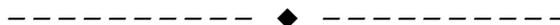

## 1 INTRODUCTION

Transmission of medical images aiming at medical consultation, diagnosis, and treatment, or for training purposes demands highly reliable and high-speed communication systems. Since, typically, the medical images contain large amounts of crucial clinical data; no visible encoding error is tolerated in the clinically important regions. Hence, advanced methodologies for gaining reconstructed image, in fairly acceptable transmission rate, make it possible to transmit a medical image over a noisy channel. However, even by utilizing broadband connections, the transmission of medical images has not been entirely successful; although using more parity bits enables a higher protection of the data against channel noise, the transmission time and the complexity of the channel coding considerably increase [1].

There are two ways in which the image information may be updated from a low-resolution mosaic of images to a high-resolution display of selected images. The first possibility is to re-transmit the entire full-resolution image to the viewing console which is a waste of bandwidth. The second option is to transmit only the difference information between the low resolution image and the full-resolution image [2].

Progressive image transmission (PIT) is a class of image transmission techniques where the information of the image is transmitted in several successive stages. The main goal of progressive transmission of images is to allow the receiver to recognize relevant features in an image as quickly as possible at minimum cost.

Wavelet transform is particularly well adapted to progressive transmission. Data compression is the process of converting data files into smaller files for efficient storage and transmission. A fundamental goal of data compression is to reduce the bit rate for transmission or storage while maintaining an acceptable fidelity or image quality.

There are two types of image compression techniques, namely i) Lossless and ii) Lossy compression. In this work lossy compression technique is used. A wavelet representation provides access to a set of data at various levels of detail. However, wavelet analysis differs from Fourier analysis such that the different signal frequencies are described by individual wavelet basis where the functions are localized rather than global.

Hamming code is a forward error control code which is well known for its single-bit error detection and correction capability. Hamming encoding and decoding are easy to be implemented with less time consuming in programming.

A method of progressive image transmission by using the 3 level haar wavelet transform and Hamming code algorithm is presented in this paper. The complete system is coded in VHDL.

The paper is organized into five sections. Section II gives the principle of wavelet transform to compress the image. Section III describes the Hamming code, the encoding and decoding procedure of Hamming code. Section IV gives example, experimental results and their discussions. The concluding remarks are presented in section V.

## 2 PRINCIPLE OF WAVELET TRANSFORM

The Wavelet analysis is a technique to transform an array of N numbers from their actual numerical values to an array of N wavelet coefficients. Each wavelet coefficient

________________________

- *P. Jagatheeswari is with the Department of Electrical and Electronics Engineering, Udaya School of Engineering, Nagercoil, Tamilnadu, India and Research Scholar, Anna University, Coimbatore, India.*
- *Dr. M. Rajaram is Professor and Head, Department of Electrical and Electronics Engineering, Government College of Engineering, Tirunelveli, Tamilnadu, India.*

represents the closeness of the fit (or correlation) between the wavelet function at a particular size and a particular location within the data array. By varying the size of the wavelet function (usually in powers-of-two) and shifting the wavelet so that it covers the entire array, one can build up a picture of the overall match between the wavelet function and the data array. Wavelet transform decomposes an image into a set of different resolution sub-images, corresponding to the various frequency bands. Wavelets are a class of functions used to localize a given signal in both space and scaling domains. Wavelets automatically adapt to both the high-frequency and the low frequency components of a signal by different sizes of windows [3]. Wavelets are functions generated from one single function ψ (as shown in the following equation), which is called mother wavelet, by dilations (a) and translations (b).

$$\psi_{a,b}(x) = |a|^{-\frac{1}{2}} \psi(\frac{x-b}{a}) \quad (1)$$

Where ψ must satisfy the following conditions.

$$\int_{-\infty}^{\infty} \psi(x) dx = 0 \quad (2)$$

and

$$\int_{-\infty}^{\infty} |\psi(x)|^2 dx = 1 \quad (3)$$

Wavelet transform is the representation of any arbitrary signal x(t) as a decomposition of the wavelet basis or write x(t) as an integral over a and b of $\psi_{a,b}$. In this work Discrete Wavelet Transform (DWT) is used. It is the discretized version of the continuous wavelet transforms (as defined by equation 1), for efficient computer implementation. DWT of signal x(t) is defined by the equation:

$$x(t) = \sum_{m,n} C_{m,n} \psi_{m,n}(t) \quad (4)$$

Where,

$$C_{m,n} = 2^{\frac{-m}{2}} \int_{-\infty}^{\infty} x(t) \psi_{m,n}(t) dt \quad (5)$$

The coefficients $C_{m,n}$ characterizes the projection of x(t) onto the base formed by $\psi_{m,n}(t)$. DWT is implemented using the subband coding method. The whole subband process consists of a filter bank (a series of filters), and filters of different cut-off frequencies, used to analyze the signal at different scales. The procedure starts by passing the signal through a half band high-pass filter and a half band low-pass filter. The filtered signal is then down-sampled. Then the resultant signal is processed in the same way as above. This process will produce sets of wavelet transform coefficients that can be used to reconstruct the signal.

### 2.1 Haar wavelet transform

Haar wavelet is the simplest wavelet. Haar transform or Haar wavelet transform has been used as an earliest example for orthonormal wavelet transform with compact support. The Haar wavelet transform is the first known wavelet and was proposed in 1909 by Alfred Haar. The Haar function can be described as a step function ψ(x) as follows:

$$\psi(x) = \begin{cases} 1 & 0 \leq x \leq 0.5 \\ -1 & 0.5 \leq x \leq 1 \\ 0 & otherwise \end{cases} \quad (6)$$

This is also called mother wavelet. In order to perform wavelet transform, Haar wavelet uses translations and dilations of the function, i.e. the transform make use of following function

$$\psi(x) = \Psi(2^j x - k) \quad (7)$$

Translation / shifting $\psi(x) = \psi(x - k)$

Dilation / scaling $\psi(x) = \psi(2^j x)$

where this is the basic works for wavelet expansion.

With the dilation and translation process as in Eq.(7), one can easily obtain father wavelet, daughter wavelet, granddaughter wavelet and so on [9].

### 2.2 Properties of Haar wavelet transform

The properties of the haar transform are described as follows:

1. Haar Transform is real and orthogonal.
   Hr = Hr* (8)
   Hr $^{-1}$ = Hr $^r$ (9)
   Therefore Haar transform is a very fast transform.
2. The basis vectors of the Haar matrix are ordered sequentially.
3. Haar Transform has poor energy compaction for images.
4. Orthogonality: The original signal is sp1it into a low and a high frequency part and filters enabling the splitting without duplicating information are said to be orthogonal.
5. Linear Phase: To obtain linear phase, symmetric filters would have to be used.
6. Compact support: The magnitude response of the filter should be exactly zero outside the frequency range covered by the transform. If this property is satisfied, the transform is energy invariant.
7. Perfect reconstruction: If the input signal is transformed and inversely transformed using a set of weighted basis functions and the reproduced sample values are identical to those of the input signal. The transform is said to have the perfect reconstruction property. If, in addition no information redundancy is present in the sampled signal, the wavelet transform is, as stated above, orthonormal [5].

### 2.3 The advantages of Haar Wavelet transform

- Best performance in terms of comu1ation time.
- Computation speed is high.
- Simplicity
- HWT is efficient compression method.
- It is memory efficient, since it can be calculated in place without a temporary array.

### 2.4 Procedure for Haar wavelet transform

To calculate the Haar transform of an array of $n$ samples:

1. Find the average of each pair of samples. ($n/2$ averages)
2. Find the difference between each average and the samples it was calculated from. ($n/2$ differences)
3. Fill the first half of the array with averages.
4. Fill the second half of the array with differences.
5. Repeat the process on the first half of the array. (The array length should be a power of two)

## 3 HAMMING CODE

In the transmission of images over a noisy channel using transform source coding, reconstructed image quality is substantially degraded by channel errors. As a result, for noisy channel applications it is necessary to correct the channel errors. Nowadays, usually data is transmitted through a communication channel in the form of binary codes, which undesirably their data bits can be masked and changed by noise. Detecting and correcting these errors is important. There are various types of interferences in transmission channel, which can negatively impact data transmission between the transmitting station and the receiving station, and both random and burst errors have occurred during transmission [8]. Since, typically, medical images contain large amounts of crucial clinical data; no visible encoding error is tolerated in the clinically important regions. Hamming code is an efficient and simple coding technique that can be used for error control in image transmission.

A Hamming code is an error-correcting code named after its inventor, Richard Hamming. Hamming code is a forward error correction code which works by adding check bits to the outgoing data stream. Adding more check bits reduces the amount of available bandwidth, but also enables the receiver to correct for more errors. Hamming codes can detect up to two simultaneous bit errors, and correct single-bit errors; thus, reliable communication is possible when the Hamming distance between the transmitted and received bit patterns is less than or equal to one. By contrast, the simple **parity** code cannot correct errors, and can only detect an odd number of errors.

The (n, k, t) Hamming code refers to an 'n' bit code word having 'k' data bits (where n>k) and 'r' (=n-k) error-control bits called redundant or redundancy bits with the code having the capability of correcting 't' bits in the error (i.e., 't' corrupted bits).

If the total number of bits in a transmittable unit (i.e., code word) is 'n' (=k+r), 'r' must be able to indicate at least 'n+l' (=k+r+1) different states. Of these, one state means no error, and 'n' states indicate the location of an error in each of the 'n' positions. So 'n+l' states must be discoverable by 'r' bits; and 'r' bits can indicate $2^r$ different states. Therefore, $2^r$ must be equal to or greater than 'n+l': $2^r \geq n+1$ or $2^r \geq k+r+l$.

The value of 'r' can be determined by substituting the value of 'k' (the original length of the data to be transmitted). For example, if the value of 'k' is '7,' the smallest 'r' value that can satisfy this constraint is '4'.

The following general algorithm generates a single-error correcting Hamming code for any number of bits.

1. Number the bits starting from 1: bit 1, 2, 3, 4, 5, etc.
2. Write the bit numbers in binary. 1, 10, 11, 100, 101, etc.
3. All bit positions that are powers of two (have only one 1 bit in the binary form of their position) are parity bits.
4. All other bit positions, with two or more 1 bits in the binary form of their position, are data bits.
5. Each data bit is included in a unique set of 2 or more parity bits, as determined by the binary form of its bit position.
    i. Parity bit 1 covers all bit positions which have the least significant bit set: bit 1 (the parity bit itself), 3, 5, 7, 9, etc.
    ii. Parity bit 2 covers all bit positions which have the second least significant bit set: bit 2 (the parity bit itself), 3, 6, 7, 10, 11, etc.
    iii. Parity bit 4 covers all bit positions which have the third least significant bit set: bits 4–7, 12–15, 20–23, etc.
    iv. Parity bit 8 covers all bit positions which have the fourth least significant bit set: bits 8–15, 24–31, 40–47, etc.
    v. In general each parity bit covers all bits where the binary AND of the parity position and the bit position is non-zero.

The form of the parity is irrelevant. Even parity is simpler from the perspective of theoretical mathematics, but there is no difference in practice.

The key thing about Hamming Codes that can be seen from visual inspection is that any given bit is included in a unique set of parity bits. To check for errors, check all of the parity bits. The pattern of errors, called the error syndrome identifies the bit in error. If all parity bits are correct, there is no error. Otherwise, the sum of the positions of the erroneous parity bits identifies the erroneous bit. For example, if the parity bits in positions 1, 2 and 8 indicate an error, then bit 1+2+8=11 is in error. If only one parity bit indicates an error, the parity bit itself is in error.

## 4 EXPERIMENTAL RESULTS AND DISCUSSION

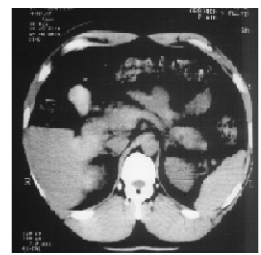 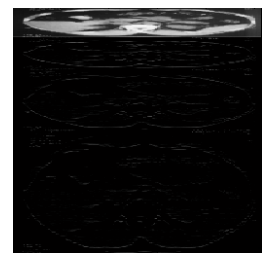

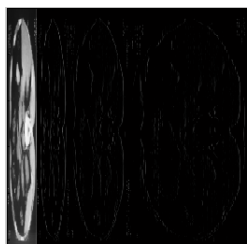
Fig. 1. Original Abdomen Image

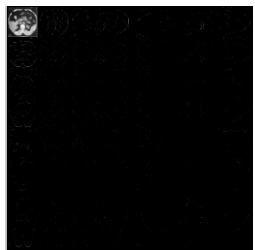
Fig. 2. Row-wise compressed image

Fig. 3. Column-wise compressed image

Fig. 4. 3-level Haar wavelet compressed image

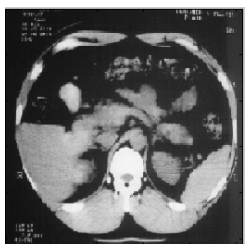
Fig. 5. Decompressed Image

Fig 1. is the original CT abdomen image obtained from Kanyakumari government medical college, Asaripallam, Nagercoil. The Haar wavelet transform is coded in VHDL and the row-wise and column-wise compressed image is shown in fig 2. and fig 3. respectively. Fig 4. shows the overall compressed image when 3-level Haar wavelet is applied. This compressed image is coded using Hamming code for error control and transmitted. In the receiver section, error correction (if any) is done by decoding. Then the inverse 3-level Haar wavelet transform process is applied to get the decompressed image as shown in fig 5.

## 4 CONCLUSION

This paper reported is aimed at developing computationally efficient and effective algorithm for lossy image compression using wavelet techniques. So this proposed algorithm developed to compress the image so fastly. The promising results obtained concerning reconstructed image quality as well as preservation of significant image details, while on the other hand achieving high compression rates.

Future works on this field may include implementing finite state vector quantization (FSVQ) with combination of wavelet transform, which may give further improvements in image quality at high compression ratios.

## ACKNOWLEDGMENT

Special Thanks Dr. S. N. Sivanandam and Dr. R. Dhanasekaran the members of Author's Ph.D supervisory committee.

## REFERENCES

[1] Jen-Lung Lo, Saied Sanei and kianoush Nazarpour " An adaptive source channel coding with feedback for progressive transmission of medical images ", International Journal of tele-medicine and applications Volume 2009, Article ID 519417, 12 pages

[2] I.Pitas, Digital Image Processing Algorithms and Applications . New York : Wiley 2000.

[3] L.Prasad and S.Iyenqar, Wavelet Analysis with Applications to Image processing. CRC press,1997.

[4] Jayantha Kumar Debnath, Newaz Muhammad Syfur Rahim and Wai –Keung Fung, "A modified vector quanntization based image compression technique using wavelet transform", IEEE proceedings of the International Joint conference on neural Networks, pp.171 – 176, 2008

[5] P.Raviraj and M.Y. Sanavullah," The modified 2D-Haar Wavelet transformation in Image compression", Middle-East Journal of scientific Research 2(2):73-78, 2007.

[6] Xiao-Yan Xu, Philip Chen and Juandai, " Hybrid encoding analysis of fractal Image compression method based on wavelet transform", IEEE proceedings of the seventh International conference on Machine learning and cybernetics, Kunming, 12-15. July 2008.

[7] J.LU, A. Nosrantinia and B.Aazhang, "Progressive joint source channel coding in feedback channels", Proceedings of Data Compression conference (DCC '99), pp.140-148, Snowbird, utah, USA ,March 1999.

[8] Etzion, Tuvi, "Constructions for perfect 2-burst-correcting codes", IEEE Transactions on Information Theory, Vol. 47, pp. 2553-2555, 2001

[9] Phang Chang, Phang Piau, "Haar wavelet matrices designation in numerical solution of ordinary differential equations", IAENG International Journal of Applied Mathematics, 38:3, IJAM_38_3_11.

[10] R.W.Hamming, "Error Detection and Error Correction codes", Bell System Tech. Journal, Vol. 29, pp 147-160, April 1950.

[11] S.Lin, An introduction to error-correcting codes, Englewood Cliffs, N.J.: Prentice Hall, 1970

**P. Jagatheeswari** received her B.E. degree in Electrical and Electronics Engineering (2002), and M.E. degree in Process Control and Instrumentation (2004) in Anamalai University, India. Currently she is pursuing her Ph.D degree in Anna University Coimbatore, India. From 2003 she is working as Lecturer in the department of Electrical and Electronics Engineering at Udaya School of Engineering, Nagercoil, India. She is a member of ISEEE (Indian Society of Electrical and Electronics Engineers). Her current research intrests include image enhancement and channel coding.

**Dr. M. Rajaram** received his B.E. degree in Electrical and Electronics Engineering (1981) from Madurai University, M.E. (1988) and Ph.D (2004) degree from Bharathiyar University, Coimbatore, India. He is having the teaching experience of 28 years and his research interests are computer Science and Engineering, Electrical Engineering and power Electronics.He is the author of over 120 publications in various International and National Journals. 5 Ph.D scholars and M.S(By Research) scholars have been awarded under his supervision. At present he is supervising 15 Ph.D Scholars.